# Beyond Chemical 1D knowledge using Transformers

Ruud van Deursen[1], Igor V. Tetko[2,3] and Guillaume Godin[1]

[1]Firmenich International SA, Rue de la Bergère 7, 1242 Satigny, Switzerland,
[2]Helmholtz Zentrum München - Research Center for Environmental Health (GmbH), Institute of Structural Biology, Ingolstädter Landstraße 1, D-85764 Neuherberg, Germany,
[3]BigChem GmbH, Valerystr. 49, D-85716, Unterschleißheim, Germany, itetko@bigchem.de
Corresponding author: guillaume.godin@firmenich.com

## Abstract

In the present paper we evaluated efficiency of the recent Transformer-CNN[1] models to predict target properties based on the augmented stereochemical SMILES. We selected a well-known Cliff activity dataset[2] as well as a Dipole moment dataset[3] and compared the effect of three representations for R/S stereochemistry in SMILES.[4] The considered representations were SMILES without stereochemistry (noChiSMI), classical relative stereochemistry encoding (RelChiSMI) and an alternative version with absolute stereochemistry encoding (AbsChiSMI). The inclusion of R/S in SMILES representation allowed simplify the assignment of the respective information based on SMILES representation, but did not always show advantages on regression or classification tasks. Interestingly, we did not see degradation of the performance of Transformer-CNN models when the stereochemical information was not present in SMILES. Moreover, these models showed higher or similar performance compared to descriptor-based models based on 3D structures. These observations are an important step in NLP modeling of 3D chemical tasks. An open challenge remains whether Transformer-CNN can efficiently embed 3D knowledge from SMILES input and whether a better representation could further increase the accuracy of this approach.

## Introduction

In the 1980s, the SMILES (simplified molecular-input line-entry system) was introduced as a text representation for the depth-first tree-traversal of a molecular graph.[4] A SMILES string thus intrinsically encodes the internal connectivity of the vertices without explicitly considering the vertex index in the graph. Such representation is important since for any given molecule, multiple valid SMILES can be written with simple alteration of the atoms order. The possibility to write multiple SMILES for the same molecule was used for data augmentation in predictive and generative models.[5] In SMILES, atom properties can be explicitly added to an atom. Apart from properties such as isotopes, formal charges and number of hydrogens, Cahn-Ingold-Prelog stereochemistry for tetrahedral stereocenters is typically defined using the annotations "@@" for clockwise and "@" for anti-clockwise chirality (RelChiSMI). Contrary to the convention that the atom order in a SMILES string is irrelevant, the RelChiSMI annotation used for stereochemistry is strictly depending on the order of the R-groups in the molecule (Figure 1). Indeed, swapping the index of R-groups may result in a change of clockwise to anti-clockwise chirality on the stereocenter, resulting in a change of the RelChiSMI between the chiral tags @@ and @. Consequently, machine learning methods need to learn to identify distant patterns in molecules to understand the stereochemistry for any stereo center in the molecule. We may thus expect that a model needs a vast amount of data to correctly learn stereochemistry using the proposed encoding. For many tasks in which stereochemistry is important, there are, however, typically rather limited data.

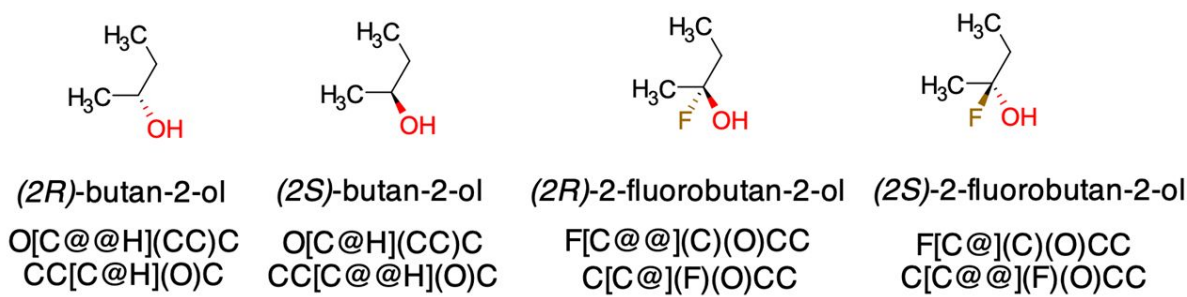

**Figure 1:** Examples showing the current SMILES convention @/@@-annotation for tetrahedral stereochemistry. From left to right: *(2R)*-butan-2-ol, *(2S)*-butan-2ol, *(2R)*-2-fluorobutan-2-ol and *(2S)*-2-fluorobutan-2-ol. For each molecule we have displayed two possible SMILES strings. Note, the @/@@-syntax changes with permutations to the order of the R-groups.

In the current RelChiSMI format, the interpretation of the stereochemistry is dependent on seven combinations, i.e., the relative positions between the 4 R-groups (six combinations) plus the sign "@@" and "@" for clockwise and counterclockwise, respectively. The number of combinations is larger than the number of possible stereochemical solutions, which are two (**R** or **S**) or three (**R, S** or ***racemic. i.e., stereochemistry is not defined***). The current representation is thus overdetermined and has a larger number of possible combinations rather than a number of target configurations. Therefore, we hypothesize that machine learning may need a vast amount of data to correctly understand stereochemistry.

Herein we analyze the performance of models to understand the R/S stereochemistry based on several SMILES conventions. Following DeepSmiles[6] spirit, we proposed a new convention for the writing of absolute stereochemistry in SMILES strings. We called this convention ***AbsChiSMI*** for "absolute chiral SMILES". In this convention the priority of the R-groups is always considered in clockwise order (a) - (b) - (c) with an indication if the group with highest priority (a) points forward (UP; *R*), points backward (DOWN; *S*) or is racemic (RACEMIC *R/S* - either not determined or ? are possible). These absolute forms are flagged immediately after the stereocenter using the characters "^" (UP), "_" (DOWN) or absent (RACEMIC), respectively (Figure 2). In the convention the character used does not change with the relative position of the R-groups in the SMILES syntax. Based on the written form a chemist can rapidly draw one or multiple correct images for the chiral center of the molecule without having to know the exact position of the R-group in the tetrahedron (Figure 1). As displayed in the figure, the liberty to position the groups in the graphical drawing is at the chemist's discretion. Based on the provided character, I.e. UP, DOWN or RAC, the chemist only needs to draw the correct bond to satisfy the *rac-, R- or S*-form of the molecule. Additionally, the stereochemistry is defined as a local property of the stereocenter in line with naming conventions of the molecules *(2R)*-butan-2-ol or *(2S)*-butan-2-ol. Finally, we also investigated a blank RACEMIC SMILES representation, named NoChiSMI. In NoChiSMI, all chiral tags with the @/@@ annotations were removed from the SMILES string rendering all stereocenters as "undefined".

Herein we present the results for Transformer-CNN models using these three formats, on the endpoints Cliff Activity[2] and dipole moment[3] and radius of gyration, for which stereochemistry is considered as an important property to explain biological activity and dipole of molecules, respectively.

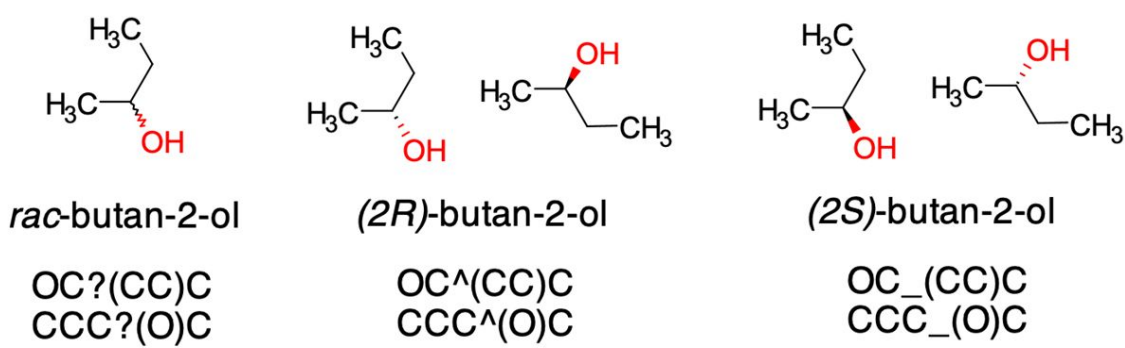

**Figure 2:** Examples showing the AbsChiSMI syntax. The examples are the molecules rac-butan-2ol, (2R)-butan-2-ol and (2S)-butan-2-ol and are displayed with the possible SMILES strings defining the molecules.

## Methods

### Library generation

A combinatorially exhaustive library of 47,502 unique molecules were enumerated *in silico* using a text-replacement procedure. The templates [C@@H](R1)(R2)R3 and [C@H](R1)(R2)R3 were used for molecules with three different non-hydrogen R-groups. The templates [C@@](R1)(R2)(R3)R4 and [C@](R1)(R2)(R3)R4 for molecules with four different non-hydrogen R-groups. For the library creation we have used 28 different R-groups. All molecules were created using 3 or 4 different R-groups, leading to a balanced dataset of *R-* and *S*-compounds only. All molecules were subsequently converted to canonical SMILES using RDKit[11] v03.2020 and annotated with the correct label for absolute stereochemistry *R* or *S*. The SMILES were subsequently augmented using random SMILES augmentation. The first SMILES code in the augmentation was always the canonical SMILES for the molecule.

## Baseline Model to Predict R/S target

We used exactly the same dataset to compare the effect of different representations and converted it to AbsChiSMI or RelChiSMI SMILES. These SMILES were used to generate fingerprint vectors composed of character frequency. For the baseline model we used a neural network with one hidden layer (ReLU) followed by a Sigmoid layer to predict the stereochemistry class R or S of the molecules.

# Results

If RelChiSMI representation was used, one symbol "@" was used to encode both S and R enantiomers. The frequency of this symbol was not sufficient to discriminate between S and R enantiomers and the classifier failed to classify both types of molecules and just predicted all of them as S enantiomers, which was the class with the largest number of samples. This result is expected, as the stereochemistry type depends on the way of writing the SMILES. Indeed, the @-characters define the equivalency of clockwise and counter-clockwise rotation respectively. The classification of the stereocenter as R or S still depends on the order of priorities for the R-groups of the stereocenter. Vise versa, in AbsChiSMI separate symbols were used for each of the enantiomers that trivially resulted in 100% of accuracy.

This simple model was used just to illustrate that simplification of the representation of stereochemistry could be beneficial for the interpretation of data and machine learning.

## Results on stereochemical related properties

We have also investigated the influence of the SMILES versions on two datasets where stereochemistry can play a significant role. One dataset was a regression task on dipole moments of 10,071 molecules.[3] Another dataset was the radius of gyration shared by Ella Gale (from unpublished work). Last one was classification task to distinguish between classes of compounds that can introduce large or negligible changes of activity of molecules (activity cliffs) due to the change of their stereochemistry.[2] To our knowledge, this is the first time we introduce a classification method based on NLP for cliff activity. The dataset included 3,838 molecules, was cute like this: 1792 cliff & 2036 non cliff.. We have selected these datasets since stereochemistry is less relevant for datasets with physico-chemical properties, which are typically used in benchmarking studies.

We tested several existing descriptors in combination with advanced machine learning architectures, namely Deep Learning Neural Networks (DNN),[7] Random Forest (RF),[8] XGBOOST,[9] DEEPCHEM text Convolutional Neural Network (TEXTCNN),[10] Associative Neural Network (ASNN),[11] Convolutional Neural Fingerprint (CNF)[5,12] and Least Squares SVM (LS-SVM)[13] available in OCHEM.[14] Each machine learning method was used to develop a model with 20 types of descriptors available in OCHEM, which included both 2D and 3D descriptors. The conversion of structures to 3D structures from the SMILES was done using the Corina program.[15]

We also developed three NLP models, based on transformer-CNN architecture[16] for RelChiSMI, AbsChiSMI and NoChiSMI formats. First we developed autoencoder models based on ChEMBL.[17] To do this we converted SMILES to each of the above formats, augmented data using random SMILES and trained Transformer as described in ref[1]. Then we used these models to analyze both datasets using exactly the same parameters.

For descriptor-based approaches we selected 3-5 models with the highest accuracy and calculated a consensus model, which was compared with the performance of the Transformer-CNN models. The results for models with the highest accuracy for individual descriptor-based models as well as for the consensus models are reported in Table 1.

**Table 1.** Performance of different methods for analyzed regression and classification tasks. The higher values of statistical parameters correspond to models with better prediction powers.

| Model\Targets | Cliff Activity | | Dipole Moment | | Radius of Gyration | |
|---|---|---|---|---|---|---|
| Metric | Area under the curve (AUC) | | Coefficient of determination ($r^2$) | | | |
| Augmentation type | X1 or best | X10 or ensemble | X1 or best | X10 or ensemble | X1 or best | X10 or ensemble |
| Trans-CNN Rel | 0.81 | **0.84** | 0.58 | **0.65** | 0.82 | 0.89 |
| Trans-CNN No | 0.8 | **0.84** | 0.58 | **0.65** | 0.84 | 0.89 |
| Trans-CNN Abs | 0.81 | 0.83 | 0.57 | 0.61 | 0.84 | **0.9** |
| GIN | 0.72 | *NA* | 0.47 | *NA* | 0.83 | *NA* |
| DNN | 0.79 | 0.82 | 0.62 | **0.67** | **0.92** | 0.93 |
| XGBoost | 0.82 | **0.86** | 0.62 | 0.66 | 0.91 | 0.92 |
| RF | **0.84** | **0.86** | 0.62 | 0.64 | 0.92 | **0.93** |
| TextCNN | NA | 0.82 | NA | 0.62 | 0.73 | 0.83 |
| TextCNF | 0.65 | 0.79 | 0.51 | 0.57 | 0.33 | 0.81 |
| ASNN | 0.79 | **0.84** | 0.6 | **0.67** | 0.91 | 0.92 |
| LS-SVM | **0.85** | **0.86** | **0.64** | 0.66 | **0.92** | **0.93** |

NA – not available. Trans-CNN – Transformer Convolutional Neural Network,[1] GIN – Graph Isomorphism Network,[18] DNN – Deep Learning Neural network,[7] XGBoost,[9] RF – Random Forest,[9] TEXTCNN - DEEPCHEM Text Convolutional Neural Network,[10] CNF - Convolutional Neural Fingerprint (CNF),[5,12] ASNN - Associative Neural Network,[11] LS-SVM – Least Squares Support Vector Machines.[13] The abbreviations Rel, No and Abs define the usage of the formats RelChiSMI, NoChiSMI and AbsChiSMI in the Transformer-CNN, respectively. Best: best model performance other 24 descriptors (including CDK, RDkit, Dragon, AlvaDesc which are almost always the best one).

The use of augmentation provided a significant increase of the accuracy of the Transformer-CNN, especially for the regression task. This method got slightly better accuracy than, e.g., the best individual DNN model as well as the consensus DNN model for Cliff Activity classification task. For the regression task to predict dipole moment the difference was smaller but still in favor of the Transformer. Random Forest and LS-SVM produced slightly better results than Transformer but only for the consensus models. The statistical parameters for these models were sometimes 0.01 higher compared to the Transformer models that could be considered as marginal. It should be mentioned that the LS-SVM method also encapsulated a hyperparameter optimization trick during the LS projection, while most of other algorithms were used only with the default set of hyperparameters. This difference may explain the similar performance of this approach in comparison to Transformer-CNN. The CDDD[19] encoding did not provide better results than Transformer-CNN. Only the combination of LS-SVM and CDDD resulted in a model, which had accuracy comparable to that of Transformer-CNN. For the radius of gyration regression task, AbsChiSMI gave very slightly better results than RelChiSMI or NoChiSMI. For this task, the LSSVM and ASNN ensemble or best models are slightly better than the Transformer-CNN models. This is the first time that we saw an interest to use the new Absolute format.

The different SMILES encoding (RelChiSMI, AbsChiSMI and NoChiSMI) provided similar results for the classification task. For the regression task the AbsChiSMI representation provided a model can be either slightly better or worse performance as compared to RelChiSMI /NoChiSMI ones.

We also tested the GIN[18] graph model to see if the graph methodology could provide similar accuracy, but the performance of this method was significantly lower. In GIN, the atom feature vector includes the R/S information as many other atomic properties. It was reported that graph models performed better than classical descriptors in classical regression benchmarks in single task and multi task. The graph isomorphic network is a continuous message passing process around each atomic layer of a molecule depending of the deep of the network.[20] It has been reported that GIN is a generalized CNN on the graph space.[21] However, it does not mean that a graph can substitute a CNN in the 3D shape of a molecule. In our experiment, we clearly saw that this method did not calculate a good performance for Cliff activity task (AUC: 0.72).

Considering that individual models developed using 3D descriptors provided on average lower performance or similar than Transformer-CNN developed with NoChiSMI, we can conclude that the Transformer embedding can capture 3D information even if this information is not directly available in the SMILES. Alternately, the relationship between stereochemistry and the endpoint may not be readily understood by the model or may play a subordinate role to achieve the best predictions.

## Discussion

The advantage of the new AbsChiSMI format can be clearly seen to predict the absolute R- and S-stereochemistry in synthetic baseline model, since "@" representation used in traditional RelChiSMI does not allow to easily infer the enantiomer type of molecules. The classification task using AbsChiSMI was trivial, since this new format directly encoded R/S using "^" and "_"characters, respectively. The absolute stereochemistry could also improve models for endpoints for which stereochemistry is expected to play a role. Such endpoints may include prediction of olfaction and or taste sensation. Example of such difference is the pair of *(R)*-limonene and *(S)*-limonene with a citrus and mint scent, respectively. It is also well known that stereochemistry plays an important role for drugs where one enantiomer cay save life while the other may cause significant harm.[22]

Interestingly, we have not seen significant differences between when using AbsChiSMI, RelChiSMI and NoChiSMI for dipole moment, radius of gyration or Cliff Activity predictions . We think that this result demonstrates the power of Transformer-CNN to infer some 3D information directly from SMILES that do not contain stereochemical information. This phenomenon is however not yet clearly understood and will be further investigated. The possibility that Transformer-CNN does not require explicit information may indicate that the used embedding layer may account for a chemically and physically realistic representation of 3D in chemistry. It remains possible that stereochemistry was not so important for both analyzed tasks, but we are not in favor of this hypothesis. Finally, we have found that AbsChiSMI can improve very slightly accuracy for the radius of gyration regression task, we expect to find other applications of this new format based on our synthetic results.

# Conclusions

We have demonstrated that AbsChiSMI could allow direct embedding information about the R/S and thus making classification of molecules for this property a trivial task. We did not see a clear advantage of AbsChiSMI on several other investigated tasks for which stereochemistry was considered to be a relevant one, except for the radius of gyration case. Interestingly, we did not see either an advantage to use RelChiSMI versus NoChiSMI using Transformer-CNN. Identical results of No/Rel representations may suggest that the two proposed Rel/Abs encoding was not fully efficient or/and that 3D descriptors and used conversion of SMILES to 3D conformations possibly did not efficiently capture the stereochemistry. In general, Transformer-CNN showed better results than the models developed with individual descriptor sets using several popular machine learning methods such as DNN, RF, XGBoost and similar or slightly lower performance to the consensus models developed by combining best individual models, only radius of gyration task is still in favor of classical methods. In general, Transformer-CNN provided good models for analyzed tasks and demonstrated robust performance when using different SMILES representations. We also hope that the proposed AbsChiSMI SMILES representation will be useful for machine learning.

# Data and Code availability

The code to convert RelChiSMI to AbsChiSMI and the used datasets are provided as supporting information and also available at http://github.com/firmenich/AbsChiSMIxxx. The Transformer-CNN using three different SMILES representations could be accessed online at http://ochem.eu.

# Acknowledgement

Guillaume Godin wants to thank Ellen Gale for sharing the radius of gyration dataset to benchmark Transformer-CNN models.

85–100.